\documentclass[prb,twocolumn,showpacs,amsmath,amssymb,floatfix]{revtex4}
\usepackage{graphicx}

\usepackage{epstopdf}

\def\beq{\begin{equation}}
\def\eeq{\end{equation}}

\begin{document}

\title{Strong-coupling electron-phonon superconductivity in
  noncentrosymmetric quasi-one-dimensional K$_2$Cr$_3$As$_3$}

\author{Alaska Subedi} 

\affiliation{Max Planck Institute for the Structure and Dynamics of
  Matter, Luruper Chaussee 149, 22761 Hamburg, Germany}

\date{\today}

\begin{abstract}
  I study the lattice dynamics and electron-phonon coupling in
  non-centrosymmetric quasi-one-dimensional K$_2$Cr$_3$As$_3$ using
  density functional theory based first principles calculations. The
  phonon dispersions show stable phonons without any soft-mode
  behavior. They also exhibit features that point to a strong
  interaction of K atoms with the lattice. I find that the calculated
  Eliashberg spectral function shows a large enhancement around 50
  cm$^{-1}$. The phonon modes that show large coupling involve
  in-plane motions of all three species of atoms. The $\mathbf{q}$
  dependent electron-phonon coupling decreases strongly away from the
  $q_z = 0$ plane. The total electron-phonon coupling is large with a
  value of $\lambda_{\textrm{ep}} = 3.0$, which readily explains the
  experimentally observed large mass enhancement.
\end{abstract}


\maketitle

\section{Introduction}

Almost all known superconductors are conventional electron-phonon
superconductors, and the phenomenon of unconventional
superconductivity, especially in one-dimensional (1D) materials, is
relatively unexplored. Therefore, the recent report of
superconductivity that is likely of an unconventional nature in
non-centrosymmetric quasi-1D K$_2$Cr$_3$As$_3$ with a $T_c$ of 6.1 K
by Bao \textit{et al.}\ is of significant importance \cite{bao15}.

K$_2$Cr$_3$As$_3$ is structurally similar to the $M$Mo$_3$$X$$_3$ or
$M_2$Mo$_6$$X$$_6$ ($M$ = metals, $X$ = chalcogens) family of
compounds first discovered in 1980 \cite{pote80,honl80}. In addition
to K$_2$Cr$_3$As$_3$, superconductivity has also been observed in two
other isostructural compounds Rb$_2$Cr$_3$As$_3$ and
Cs$_2$Cr$_3$As$_3$ with $T_c$'s of 4.8 and 2.2 K,
respectively\cite{tang15a,tang15b,tang15c}. The one-dimensional
element in $A_2$Cr$_3$As$_3$ ($A$ = K, Rb, Cs) compounds is a
[(Cr$_3$As$_3$)$^{2-}$]$_\infty$ chain, which is arranged in a hexagonal
lattice. This chain has the structure of a double-walled
subnanotube. The inner wall is made of stacked Cr$_6$ octahedra that
are face-sharing along the $c$ axis and the outer wall is made of
similarly stacked face-sharing As$_6$ octahedra. The individual chains
are separated by metal cations that provide the compensating charge
and stabilize the structure.

The superconductivity in K$_2$Cr$_3$As$_3$ exhibits various features
that are suggestive of an unconventional nature. The superconducting
state arises out of a normal state that has a large specific heat
coefficient of 70--75 mJ K$^{-2}$ mol$^{-1}$, which indicates strong
quasiparticle mass renormalization\cite{bao15,kong15}. The upper
critical field exceeds the one-band BCS estimate of the Pauli limit by
a factor of 3--4\cite{bao15,kong15,bala15,wang15}. NMR experiments
show that spin fluctuations grow as the temperature approaches $T_c$
\cite{zhi15}. Below $T_c$, the nuclear spin-lattice relaxation rate
does not show Hebel-Slichter coherence peak that is characteristic of
an isotropic $s$-wave superconductor, and the relaxation rate
decreases with a power-law behavior suggesting the presence of
zero-energy excitations in the superconducting state. The presence of
zero-energy excitations is also seen in the temperature dependence of
the penetration depth, which decreases linearly at temperatures well
below $T_c$\cite{pang15}. A fit of the angular dependence of the
superconducting gap to the superfluid density obtained from the
penetration depth shows reasonable agreement with various $p$-, $d$-
and $f$-wave models that exhibit nodes. The temperature dependence of
the superfluid density obtained from transverse-field $\mu$SR
measurements show equally good fit to both an isotropic $s$-wave model
and a $d$-wave model with line nodes\cite{adro15}. In addition,
zero-field $\mu$SR measurements show a presence of internal magnetic
field in the superconducting state, although the magnetic field is
more than 100 times smaller than that observed in Sr$_2$RuO$_4$. Raman
scattering study of phonons with frequency greater than 100 cm$^{-1}$
also show very weak electron-phonon coupling that is too small to
account for the observed superconductivity in this
material\cite{zhan15}. In addition to these experimental observations,
various first principles and Hubbard model based theoretical studies
also find that a triplet superconducting state mediated by magnetic
fluctuations to be likely present in K$_2$Cr$_3$As$_3$
\cite{jian14,wu15a,zhou15,wu15b,zhon15,wu15c}.

The experimental results that have so far been gathered on
K$_2$Cr$_3$As$_3$ rule out an isotropic $s$-wave superconducting state
in the weak-coupling limit, but they do not necessarily imply an
unconventional 1D superconductivity mediated by magnetic
fluctuations. Indeed, as pointed out by Balakirev \textit{et al.}, the
superconductivity in K$_2$Cr$_3$As$_3$ is inconsistent with such an
unconventional superconducting state\cite{bala15}. They find that the
anisotropy of the upper critical field $\gamma_H(T) = H_{c2}^{\perp} /
H_{c2}^{\parallel}$ is modest with $\gamma_H(T_c) \approx 0.35$ at
$T_c$ and $\gamma_H(0) \approx 1.5$ at 0.7 K, which suggests a rather
three-dimensional nature of superconductivity. Even though
K$_2$Cr$_3$As$_3$ is structurally very one-dimensional, first
principles calculations show that the electronic structure is
decidedly three-dimensional with the presence of a large
three-dimensional Fermi sheet in addition to two one-dimensional
sheets\cite{jian14,wu15a}. Furthermore, the upper critial field
parallel to the chain $H_{c2}^{\parallel}$ is limited by Pauli pair
breaking and is indicative of a singlet superconductivity, whereas the
upper critical field perpendicular to the chain $H_{c2}^{\perp}$ shows
no Pauli pair breaking effects and is limited by orbital pair
breaking\cite{bala15}. In a quasi-1D superconductivity,
$H_{c2}^{\perp}$ would have been limited by orbital and Pauli pair
breaking effects yielding $H_{c2}^{\perp} < H_{c2}^{\parallel}$. More
importantly, superconductivity in K$_2$Cr$_3$As$_3$ has been observed
in samples with residual resistivity ratio ranging from 10 to
50\cite{bao15,kong15,bala15,wang15}, but the $T_c$ is essentially
insensitive to the non-magnetic impurities. This suggest that the
pairing is of $s$-wave type since scattering with non-magnetic
impurities tends to average out the superconducting gap function,
which would be detrimental to unconventional superconductors as their
gap function has different signs in different regions of the Brillouin
zone.

Motivated by these observations, I have investigated the lattice
dynamics and electron-phonon coupling in K$_2$As$_3$Cr$_3$ using first
principles density functional perturbation theory calculations. The
phonon dispersions show the structure is stable, and there is an
absence of a soft phonon behavior. Instead, they exhibit features that
point to a strong coupling of K atoms with the lattice. The Eliashberg
electron-phonon spectral function is strongly enhanced around 50
cm$^{-1}$. These phonons correspond to vibrations of all three
constituent species and the coupling cannot be attributed to an
Einstein-like rattling vibration. The phonon modes that exhibit large
coupling are characterized by in-plane motion of the atoms. The
$\mathbf{q}$ dependent electron-phonon coupling is highly
anisotropic. In particular, the coupling strongly decreases as one
moves away from the $q_z = 0$ plane. The calculated total
electron-phonon coupling $\lambda_{\textrm{ep}} = 3.0$ is large and
readily explains the experimentally observed large mass enhancement.

\section{Computational approach}

The phonon dispersions and electron-phonon couplings presented here
were obtained using density functional perturbation theory within the
generalized gradient approximation (GGA) as implemented in the Quantum
{\sc espresso} package\cite{qe}. I used the ultrasoft pseudopotentials
generated by dal Corso\cite{dalc14} and cutoffs of 50 and 500 Ry for
basis-set and charge-density expansions, respectively. A $4 \times 4
\times 8$ grid was used for the Brillouin zone integration in the
self-consistent calculations. The dynamical matrices were calculated
on a $4 \times 4 \times 8$ grid and phonon dispersions and density of
states are then obtained from Fourier interpolation. A denser $8
\times 8 \times 16$ grid and a smearing of 0.008 Ry was used in the
calculation of the electron-phonon coupling. The electronic structure
and structure relaxation calculations were also performed using the
generalized full-potential method as implemented in the {\sc wien}2k
package\cite{wien2k} as a check.

\begin{table}[h!tbp]
  \caption{\label{tab:pos} Relaxed internal parameters of
    K$_2$Cr$_3$As$_3$ using non-spin-polarized pseudopotential and
    full-potential calculations. The atomic coordinates in the
    $P\overline{6}m2$ space group are: As1 $(x, -x, 0)$, As2 $(x, -x,
    0.5)$, Cr1 $(x, -x, 0.5)$, Cr2 $(x, -x, 0)$, K1 $(x, -x, 0.5)$, K2
    $(1/3, 2/3, 0)$.  }
  \begin{ruledtabular}
    \begin{tabular}{lcccccc}
      & Wyckoff & $x$ & $x$ \\
      & position &  pseudopotential     & full-potential \\
      \hline
      As1 & 3$j$  & 0.8345 & 0.8344 \\
      As2 & 3$k$  & 0.1667 & 0.1668 \\
      Cr1 & 3$k$  & 0.9175 & 0.9167 \\
      Cr2 & 3$j$  & 0.0859 & 0.0864 \\
      K1  & 3$k$  & 0.5380 & 0.5381 
    \end{tabular}
  \end{ruledtabular}
\end{table}

I used the experimental lattice parameters $a = 9.9832$ and $c =
4.2304$ \AA\ but relaxed the internal atomic position parameters of
the $P\overline{6}m2$ space group. The results of the geometry
optimization calculations performed using both the pseudopotential and
full-potential methods are summarized in Table \ref{tab:pos}. These
are in good agreement with each other and to the experimental
values\cite{bao15,tang15a}. Structural optimization leaves the As-As
distances basically unchanged but reduces the Cr-Cr distances by 0.1
\AA\ relative to the experimental values, in accord with a previous
theoretical study\cite{wu15a}.

\section{Results and discussion}

\begin{figure}
  \includegraphics[width=\columnwidth]{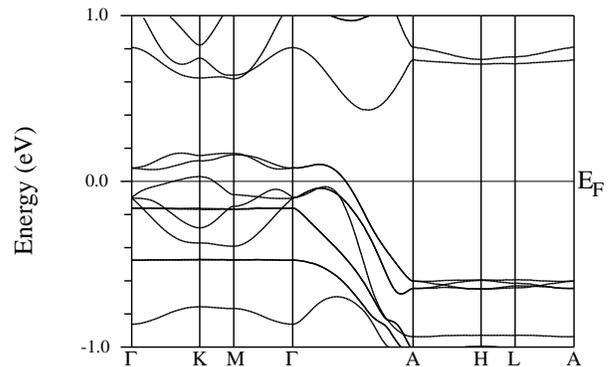}
  \caption{
    Calculated GGA band structure of non-spin-polarized
    K$_2$Cr$_3$As$_3$ in the relaxed structure. }
  \label{fig:bnd}
\end{figure}

\begin{figure}
  \includegraphics[width=0.5\columnwidth]{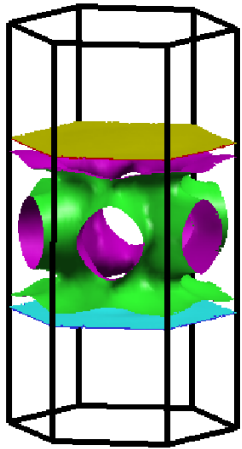}
  \caption{ (Color online) Calculated GGA Fermi surface of
    non-spin-polarized K$_2$Cr$_3$As$_3$ in the relaxed structure. }
  \label{fig:fs}
\end{figure}

The band structure and Fermi surface of K$_2$As$_3$Cr$_3$ calculated
using the relaxed internal parameters are shown in Figs.~\ref{fig:bnd}
and \ref{fig:fs}, respectively. These differ in details from the ones
presented in Refs.~\onlinecite{jian14} and \onlinecite{wu15a} that
used experimental atomic positions. The one-dimensional nature of the
material is evident from the relatively small dispersion along the
in-plane $\Gamma$--$K$--$M$--$\Gamma$ and $A$--$H$--$L$--$A$
paths. However, some bands do show moderate dispersion along these
paths, which indicates that the bonding between
[(Cr$_3$As$_3$)$^{2-}$]$_\infty$ chains is substantial. The bands
disperse strongly in the $k_z$ direction (the $\Gamma$-$A$ path), and
these reflect the short inter-atomic distances and strong bonding in
the out-of-plane direction. I find substantial As $p$ character above
the Fermi level and the presence of large amount Cr $d$ character in
the lower part of the valence band manifold, which suggests that there
is a strong covalency between Cr and As atoms. Even though the valence
bands disperse relatively little in the in-plane directions, the
valence band manifold extends from $-4.8$ to $0.2$ eV and has a band
width of 5.0 eV. This again reflects the strong covalency between Cr
and As atoms. I obtain $N(E_F) = 6.94$ eV$^{-1}$ per formula unit both
spin basis for a value of the electronic density of states at the
Fermi level. This corresponds to a calculated Sommerfeld coefficient
of $\gamma = 16.36$ mJ/mol K$^2$. The experimental $\gamma$ obtained
from specific heat capacity measurements is $\sim$70 mJ/mol
K$^2$\cite{bao15,kong15}, and this corresponds to an ehancement by a
factor of 4.3 relative to the calculated value.

There are three bands that cross the Fermi level, and these give rise
to three Fermi sheets. The two upper-lying bands cross the Fermi level
only along the $k_z$ direction and give rise to a pair of 1D
sheets. In my calculations these are degenerate, but they are
non-degenerate in the calculations that use experimental atomic
positions\cite{jian14,wu15a}. The lower-lying band crosses the Fermi
level around the $K$ point, and this gives rise to a large
three-dimensional sheet. Therefore, even though the structure of
K$_2$Cr$_3$As$_3$ is quasi-1D, the material is very three-dimensional
from the electronic structure point of view.

\begin{figure}
  \includegraphics[width=\columnwidth]{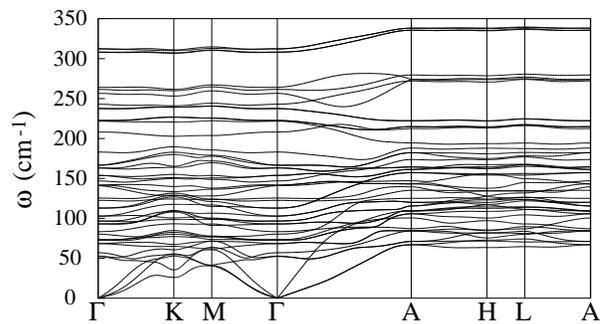}
  \caption{ Calculated GGA phonon dispersions of non-spin-polarized
    K$_2$Cr$_3$As$_3$. }
  \label{fig:pband}
\end{figure}

\begin{figure}
  \includegraphics[width=\columnwidth]{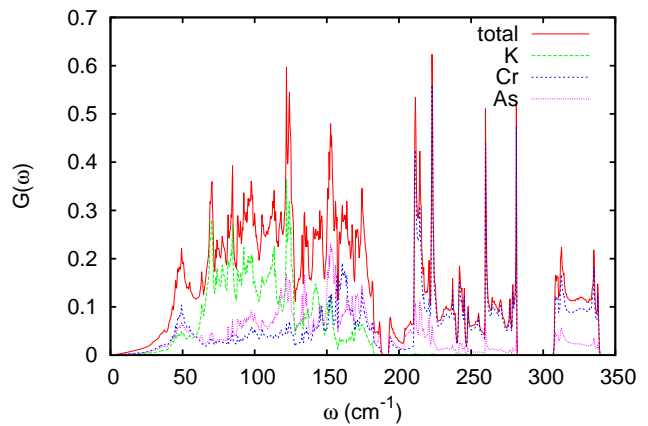}
  \caption{ (Color online) Calculated GGA phonon density of states of
    non-spin-polarized K$_2$Cr$_3$As$_3$. }
  \label{fig:pdos}
\end{figure}

The phonon dispersions and density of states (pDOS) of
K$_2$Cr$_3$As$_3$ are shown in Figs.~\ref{fig:pband} and
\ref{fig:pdos}, respectively. These again reflect the quasi-1D nature
of the material as the in-plane dispersion of the optical branches is
comparatively smaller than the out-of-plane dispersion. The strong
bonding along the $c$ axis is also seen in the large dispersion of the
out-of-plane polarized acoustic mode along the $q_z$ direction. This
branch reaches more than 100 cm$^{-1}$. Although the acoustic branches
have smaller dispersion along the in-plane direction, these do not
exhibit soft-mode behavior. This indicates weaker bonding between the
[(Cr$_3$As$_3$)$^{2-}$]$_\infty$ chains, but it also implies that the
structure is stable and not in a very close vicinity of a structural
phase transition.

The pDOS exhibits many narrow peaks that are caused by narrow band
widths of the phonon branches. The partial pDOS shows that the K atoms
contribute to phonon modes up to 200 cm$^{-1}$. Therefore, the K atoms
do not behave like a rattler that is weakly coupled to the
lattice. The interaction with K ions is also seen in the stronger
dispersion around the $K$ point for branches below 200 cm$^{-1}$. The
Cr and As atoms contribute to vibrations all across the phonon
dispersion range. In particular, both Cr and As atoms contribute to
the high-frequency phonon modes that again indicates strong covalency
between these two.

The strength of the interaction between electrons and phonons is
described in terms of the Eliashberg spectral function:
\begin{equation}
\alpha^2
F(\omega)=\frac{1}{N(E_F)}\sum_{\mathbf{k},\mathbf{q},\nu,n,m}%
\delta(\epsilon_{\mathbf{k}}^{n})\delta(\epsilon_{\mathbf{k+q}}^{m}%
)|g_{\mathbf{k},\mathbf{k+q}}^{\nu,n,m}|^{2}\delta(\omega-\omega
_{\nu\mathbf{q}}),
\label{eq:alpha}
\end{equation}
where $N(E_F)$ is the electronic density of states at the Fermi
level, $\epsilon_{\mathbf{k}}^{n}$ is the electronic energy at
wavevector $\mathbf{k}$ and band index $n$, $\omega _{\nu\mathbf{q}}$
is the energy of a phonon with wavevector $\mathbf{q}$ and branch
index $\nu$, and $g_{\mathbf{k},\mathbf{k+q}}^{\nu,n,m}$ is the matrix
element for an electron in the state $|n\mathbf{k}\rangle$ scattering
to $|m\mathbf{k+q}\rangle$ through a phonon $\omega
_{\nu\mathbf{q}}$.

\begin{figure}
  \includegraphics[width=\columnwidth]{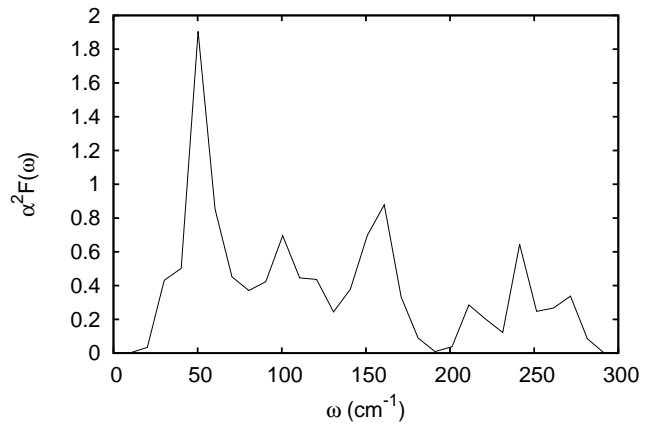}
  \caption{
    Calculated Eliashberg spectral function of K$_2$Cr$_3$As$_3$. }
  \label{fig:eliash}
\end{figure}

The calculated Eliashberg function is shown in
Fig.~\ref{fig:eliash}. It shows that phonon at all frequencies
contribute to the coupling, although the the phonon modes around 50
cm$^{-1}$ make an especially large contribution. The pDOS also shows a
peak at this frequency. Similar low-energy peaks has also been
observed in $M_2$Mo$_6X_6$ compounds, and they have been attributed to
the Einstein-like vibrations of $M$ atoms\cite{brus90,petr10}. However,
we can see from Fig.~\ref{fig:pdos} that the motion of all three
types of atoms contribute to this low-energy peak in
K$_2$Cr$_3$As$_3$. Therefore, even if these phonons have relatively
small dispersion, the coupling cannot be ascribed to an Einstein-type
rattling vibration of K atoms. I also investigated the polarization of
the phonons that show large coupling to the electrons and find that
these show strong in-plane character.

\begin{figure}
  \includegraphics[width=\columnwidth]{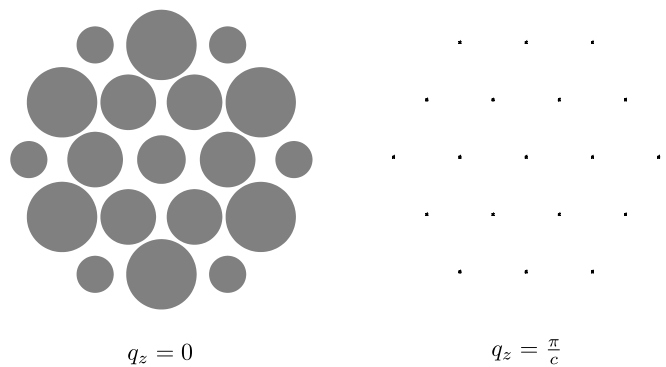}
  \caption{ Calculated $\mathbf{q}$ dependent electron-phonon coupling
    $\lambda_{\mathbf{q}}$ shown for $q_z = 0$ and $q_z = \frac{\pi}{c}$
    planes. The radius of the circle is proportional to the magnitude
    of $\lambda_{\mathbf{q}}$.}
  \label{fig:lambdaq}
\end{figure}

The $\mathbf{q}$ dependent total electron-phonon coupling
$\lambda_{\mathbf{q}}$ ($= \int_0^{\infty} \mathrm{d}\omega
\alpha^2F(\omega,\mathbf{q})/ \omega$) is plotted for the $q_z = 0$
and $\frac{\pi}{c}$ planes of the Brillouin zone in
Fig.~\ref{fig:lambdaq}. I find that the variance of the coupling is
large within a $q_z$ plane. Furthermore, the coupling strongly
decreases as one moves away from the Brillouin zone center along the
$q_z$ direction. This shows that the coupling is highly anisotropic,
and it should also result in an anisotropic superconducting gap
function. To find out which Fermi sheets contribute more to the total
electron-phonon coupling, I calculated the partial electron-phonon
couplings at several $\mathbf{q}$ points by summing over the 1D and 3D
Fermi sheets separately in Eq.~\ref{eq:alpha}. I find that the 3D
sheet makes a dominant contribution to the coupling. Therefore,
controlling the size of the 3D sheet by chemical doping, pressure or
strain might be an effective way of enhancing or suppressing
superconductivity. It will be interesting to see if magnetism or
charge density wave order appears due to nesting between 1D sheets
when the 3D sheet is suppressed.

I obtain a large value of $\lambda_{\textrm{ep}} = 3.0$ for the total
electron-phonon coupling constant $\lambda_{\textrm{ep}}
\!=\!\sum_{\mathbf{q},\nu} \lambda_{\mathbf{q},\nu}=2
\int_{0}^{\infty} \mathrm{d}\omega \alpha^2 F(\omega) / \omega$. The
logarithmically averaged phonon frequency is $\omega_{\ln} = 71$
cm$^{-1}$. These can be used to estimate the $T_c$ by using the
simplified Allen-Dynes formula, even though the use of this formula is
invalid in the present case of strong electron-phonon coupling that is
highly anisotropic\cite{carb90}. The simplified Allen-Dynes formula
has the form
\[
T_c= \frac{\omega_{\ln}}{1.2}\exp\left\{ -
\frac{1.04(1+\lambda_{\mathrm{ep}})}{\lambda_{\mathrm{ep}}-\mu^*(1+0.62\lambda_{\mathrm{ep}})}\right\}.
\]
Using a value of $\mu^* = 0.12$ for the Coulomb pseudopotential
parameter, I obtain $T_c = 17.8$ K, which strongly overestimates the
experimental $T_c$ of 6 K. A full solution of the Migdal-Eliashberg
equations, which is beyond the scope of this paper, will be required
to obtain a more accurate $T_c$ based on the calculated spectral
function. Furthermore, there might be additional reasons for the
overestimation of $T_c$. In the phonon calculations, I have neglected
the phonon anharmonicities. This can raise the phonon frequencies,
which will lower the total electron-phonon coupling. Another reason
for the overestimation might be because I have neglected the effects
of magnetic fluctuations. First principles calculations show the
presence of various magnetic interactions with in this
material\cite{jian14,wu15a}, and this will also suppress $T_c$.

The calculated total electron-phonon coupling $\lambda_{\textrm{ep}} =
3.0$ readily explains the experimentally observed large electron mass
enhancement from specific heat measurments. The electron-phonon
coupling enhances the mass by a factor of $1+\lambda_{\textrm{ep}} =
4.0$, which is very close to the experimentally observed mass
enhancement factor of 4.3. Therefore, the large mass enhancement in
this material does not come from strong electronic correlations (in
the sense of electron-electron interactions) or spin
fluctuations. This is consistent with the experimentally estimated
Wilson ration of $R_W \approx 1$\cite{kong15} that suggests a weak
magnetic enhancement. The strong coupling nature of superconductivity
in K$_2$Cr$_3$As$_3$ is also evident in the large jump of the
dimensionless specific heat $\Delta C/\gamma T_c =
2.4$\cite{bao15,kong15} at $T_c$. The strong electron-phonon coupling
superconductivity proposed here can further be confirmed from
tunneling experiments that should show a large value for the ratio
$2\Delta(0) / T_c$, where $\Delta(0)$ is the superconducting gap.

Although the calculations presented here imply that the
superconductivity in K$_2$Cr$_3$As$_3$ is caused by electron-phonon
interactions, this does not necessarily mean that the
superconductivity is dull or of the conventional BCS type. The crystal
structure of this material lacks inversion symmetry, and parity is not
a good quantum number is such a case. The pairing will likely have a
dominant $s$ character mixed with some $p$ character. This can explain
the presence of weak internal magnetic field below $T_c$ in $\mu$SR
measurements\cite{adro15}. In addition, the electron-phonon coupling
is highly anisotropic, which can lead to accidental gaps in the
superconducting state. This would be consistent with the signatures of
nodes in the gap function in the NMR\cite{zhi15} and penetration
depth\cite{pang15} measurements as well as the absence of
Hebel-Slichter peak in the NMR measurements.

The presence of such a large electron-phonon coupling in this material
provides an opportunity to study the properties of a superconducting
state in the strong coupling limit. In particular, this may help
explain why the measured upper critical fields $H_{c2}^{\perp}$ and
$H_{c2}^{\parallel}$ greatly exceeds the Pauli limit estimated for a
weakly-coupled isotropic BCS superconductor. It would be interesting
to see if the fact that the coupling strongly decreases away from the
$q_z = 0$ plane can explain why $H_{c2}^{\parallel} < H_{c2}^{\perp}$
in this material. These issues require more theoretical studies based
on the full solution of Migdal-Eliashberg equations to be fully
clarified.

As mentioned above, superconductivity is also observed in other
inorganic quasi-1D compounds including Tl$_2$Mo$_6$Se$_6$ and
Li$_{0.9}$Mo$_6$O$_{17}$, and the nature of their pairing interaction
has not been fully elucidated. The results of phonon dispersions and
electron-phonon coupling for K$_2$Cr$_3$As$_3$ presented here provides
a reference case for other quasi-1D superconductors. In particular,
the present study shows that a behavior that deviates from what is
expected from a weakly-coupled isotropic $s$-wave BCS superconductor
does not necessarily imply that the superconductivity is mediated by
magnetic interactions.

\section{Summary and conclusions}

In summary, I have investigated the lattice dynamics and
electron-phonon coupling in K$_2$Cr$_3$As$_3$ using first principles
calculations. The phonon dispersions do not show closeness to a
lattice instability and exhibits features pointing to a strong
coupling of K atoms with the lattice. The Eliashberg spectral function
is strongly enhance around 50 cm$^{-1}$, but there is coupling to high
frequency phonons as well. The peak in the spectral function
corresponds to the vibrations of all three constituent atomic species
of the material. Although the electrons do not predominantly couple to
vibrations of any specific atoms, vibrations that involve in-plane
motion of atoms show strong electron-phonon coupling. The
electron-phonon coupling shows strong $\mathbf{q}$ dependence,
especially along the out-of-plane direction. The calculated
total-electron phonon coupling $\lambda_{\mathrm{ep}}$ = 3.0 is large
and is consistent with the experimentally observed large mass
enhancement and jump in the specific heat at
$T_c$\cite{bao15,kong15}. These calculations motivate tunneling
experiments that can further confirm the strong coupling nature of
superconductivity in this material by displaying an enhanced value of
the ratio $2\Delta(0)/T_c$. The electron-phonon spectral function can
also be extracted from such experiments, and they can be compared with
the results presented here to validate the strong electron-phonon
coupling scenario proposed here.

\section{Acknowledgments}

I am grateful to David J. Singh for helpful discussions and
suggestions. This work was supported by the Swiss National
Supercomputing Centre (CSCS) under project ID s575 and by a grant from
the European Research Council (ERC-319286 QMAC).



\end{document}